\documentclass{article}

\usepackage[
  paperheight=8.5in,
  paperwidth=5.5in,
  left=10mm,
  right=10mm,
  top=20mm,
  bottom=20mm]{geometry}
\usepackage[utf8]{inputenc}

\usepackage{graphicx}
\usepackage{wrapfig}
\usepackage[bottom]{footmisc}
\usepackage{listings}
\usepackage{enumitem}

\usepackage{wrapfig}
\usepackage{ragged2e}

\usepackage{array}
\usepackage[table]{xcolor}
\usepackage{multirow}
\usepackage{booktabs}
\usepackage{hhline}
\definecolor{palegreen}{rgb}{0.6,0.98,0.6}

\usepackage{amsmath}
\usepackage{amssymb}
\usepackage{multicol}
\usepackage{lipsum}
\usepackage{hyphenat}
\PassOptionsToPackage{hyphens}{url}
\usepackage{url}

\usepackage{rotating}

\usepackage[T1]{fontenc}
\usepackage{textcomp}
\usepackage{listings}
\usepackage{setspace}

\newcommand*{\email}[1]{\small{\texttt{#1}}}

\renewcommand{\footnoterule}{%
  \kern -3pt
  \hrule width \textwidth height 0.5pt
  \kern 2pt
}

\date{}

\usepackage{titlesec}
\titleformat*{\section}{\large\bfseries}
\titleformat*{\subsection}{\normalsize\bfseries}
\titleformat*{\subsubsection}{\normalsize\bfseries}

\usepackage{biblatex}

\usepackage{subcaption}

\title{Automatic Assessment of the Design Quality of Student Python and Java Programs }

\author{
J. Walker Orr\\
Computer and Information Sciences\\
George Fox University\\
Newberg, OR 97132\\
\email{jorr@georgefox.edu}\\
}

\begin{document}
\maketitle

\begin{abstract}
Programs are a kind of communication to both computers and people, hence as students are trained to write programs they need to learn to write well-designed, readable code rather than code that simply functions correctly.
The difficulty in teaching good design practices that promote readability is the labor intensiveness of assessing student programs.
Typically assessing design quality involves a careful reading of student programs in order to give personalized feedback which naturally is time consuming for instructors.
We propose a rule-based system that assesses student programs for quality of design of and provides personalized, precise feedback on how to improve their work.
To study its effectiveness, we made the system available to students by deploying it online, allowing students to receive feedback and make corrections before turning in their assignments.
The students benefited from the system and the rate of design quality flaws dropped 47.84\% on average over 4 different assignments, 2 in Python and 2 in Java, in comparison to the previous 2 to 3 years of student submissions.
\end{abstract}

\section{Introduction}

Recently there has been increasing interest in intelligent tutoring systems for programming instruction. 
Primarily the focus has been on assisting students by giving them helpful hints on how to complete their programming assignments.
There has been a lot of work surrounding the \textit{Hour of Code} \cite{hoc}, a massively open online course intended to teach children how to program with a visual programming language.
The visual language contain constructs analogous to conditional statements but lacks \textit{loops}.
Nevertheless, the project is popular and there has been significant research into producing a intelligent tutors for it.
Typically the goal of these systems \cite{conthint, rlhint} is to suggest edits to a student's program to get it closer to correct functionality.
Another approach create an embedding for student programs that could be used to suggest hints on how to correct their programs \cite{progembed}.

However, becoming a capable and competent programmer involves more than simply writing functional code, solutions must also be readable and understandable, following commonly accepted conventions.
Even if a program works correctly when first created, the reality of most software systems is that of continuous maintenance and development.
Unnecessary complexity, convoluted logic, and unconventional style will result is mistakes in the ongoing maintenance and development of software.
Naturally, a major difficulty of teaching the practice of writing readable, well-design programs is the time-intensiveness of giving quality feedback.
Determining if a student's program works correctly can usually be assessed with unit or system integration tests.
However, providing feedback on the quality of program design generally speaking requires as close reading of the code which of course is time consuming.

In the professional setting, ``linters'' are commonly used to check for design errors and violations of programming conventions.
For example, Pylint \cite{pylint} is a commonly used ``linter'' for the Python programming language which does catch design flaws and bad practices such as the use of global variables.
Additionally, there is the \textit{PEP8} standard 
and eponymous program that checks Python programs for violations of the standards.
For Java, tools such as CheckStyle \cite{checkstyle} applies configurable rules to ensure adherence to convention and code formatting.

However useful, the needs of an educational setting differ from the professional in a few key ways.
First of all, they are intended to be used by professional programmers who are experienced and (hopefully) have a strong sense of good design.
This means that these ``linters'' are primarily checking for a short list of less egregious violations.
Typically ``linters'' check code format and use heuristics to judge code complexity.
The notion of ``the right tool for the job'' is left up to the discretion of the professional.
Some design choices such as the use of closures may be appropriate in professional setting but are generally not a good choice for an assignment in an introductory computer science class.
Hence a ``linter'' may rightfully permit some practices and design choices that are simply not appropriate for students.
Another example is the use of recursion.
Recursion is elegant tool in some settings but novice students can abuse it, for example, by recursively calling a ``main'' function to repeat the logic of a program rather than using a \textit{while} loop.
A ``linter'' typically would not even check for the use of recursion, but this bad practice can be common in introductory computer science courses.
Students, especially early on, require tighter constraints to guide them to use simple and direct approaches to relatively simple problems.
These types of constraints are not generally considered in a professional setting.
These kinds of very specific design requirements generally do not occur in a professional setting hence tools such as Pylint, PEP8, or CheckStyle do not have the ability to check for them.

A recent work targeting the professional setting, \textit{DeepCodeReviewer} \cite{deepcodereview} utilized a database of code reviews to suggest design quality improvements for C\# code segments.
It is a deep learning model trained on a propriety data taken from Microsoft's software version control system.
Likewise, the feedback it generates is also not particularly relevant to education since it is trained on professional's code who likely do not make the same mistakes as first-year computer science students.

Recently there has been some work using neural networks to assess the design quality of student programs \cite{myneuralnet}.
Though that system accurately predicted design scores and gave high-level feedback, there are some limitations both to that system and to any system based on machine learning.
Feedback on the program level, though shown to be useful, is not as helpful as feedback targeted at specific segments of the program such as functions or individual lines.
More detailed and specified the feedback will presumably be easier for a student to understand and remedy. 

\subsection{Our Approach and Contribution}

Our approach is to take design principles, model them by formally representing the principles as logical rules applicable to an abstract syntax tree.
The design principles were in general each expanded into multiple rules prescribing syntactic structures that are either best practices or are patterns that should be avoided.
The models were implemented in Haskell \cite{jones2003haskell} since each rule could be easily be stated in a declarative fashion.
These model are the first for the design quality of Python \& Java programs for an educational environment.
The implemented models was then hosted as a web service, allowing students to evaluate their programs at any time and as many times as they wished.
Finally, data was collected on the number of design quality errors made by students who had access to the models versus work done by students in previous years.
The data indicates a large drop in the rate of design errors made by students.

\section{Method}
The development of our intelligent tutoring system was a process that started with the articulation of our the principles we use to assess good design.
Consolidating both written policy and practices as well as our general intuition about what makes the design of a program good was perhaps the most difficult aspect of creating the system.
After these principles were realized and identified, they were contextualized for both Python and Java programs and more formally represented as logical rules for the respective syntactic structures.
Next a model was implemented for each language as a program written the functional programming language Haskell.
Though there are rules in common, Java and Python are very different syntactically so implementing a model was a more direct and effective approach.
Haskell's declarative nature and history of use for analysing programming languages made it a natural choice.
Finally the tutoring system was deployed as a web service, allowing students enrolled in our introductory computer science course to use it at anytime to validate their programs before the assignments were due.

\label{sect-pinciple}
\subsection{Models}

After articulating our design principles of our institution, the principles were contextualized and specified for both the Python and Java programming languages.
The model for these principles are rules represented in first order logic.
Each rule makes use of predicates that express proprieties of abstract syntax trees (AST).
The rules are in effect constraints on the space of ASTs, eliminating those with or without some features.

The purpose of the models is to detect design errors in a student's program.
It could be useful to detect segments of a student's program that are well designed in order to encourage students by applauding their success.
However, the purpose of this model is to point out mistakes so that they can be remedied by the student.
Also our institution's existing principles and written policy on program design focused heavily on avoiding mistakes.
Further, the existence of a design mistake is much more objective and readily identifiable than a good design quality.

With this in mind, our models are a collection of rules, applied to an AST, to identify individual mistakes exactly where they occur in a student's program.
Each rule in model will generate a mistake to be reported to the student if the condition of the rule is true.
The condition i.e. body of each rule consists of a logical combination of predicates.
Each predicate is simply a observation of the existence or nonexistence of property of the AST.
A predicate is simply a boolean function over ``objects'' which nodes in the AST which correspond to either statements or expressions in the source code.
A statement is an operation of a program, for example the assignment of a variable, the definition of a function, or a ``return'' from a function.
An expression is a computable value, such as an arithmetic operation, the comparison of two values, or the evaluation of a function.
Generally speaking, every modern programming language can be broken down into a combination of statements and expressions including Python.
Each expression or statement may be composed of more statements or expressions hence forming a sub-tree in the AST.
Altogether, in our model a predicate is a boolean function that operates over a sub-tree in the AST.
This allows for great expressive power and the ability to detect any characteristic that is grammatically represented in the AST.

\begin{figure}
    \centering
    
    \begin{subfigure}{.5\textwidth}
    \centering
    \begin{lstlisting}[language=Python]
def record_score(h_won):
   global human_score
   global comp_score

   if h_won:
      human_score += 1
   else:
      comp_score += 1
    \end{lstlisting}
    \caption{Global variables example.}
    \label{fig:global-example}
    \end{subfigure}%
    \begin{subfigure}{.5\textwidth}
    \centering
    \begin{lstlisting}[language=Python]
def find(my_list, value):
   i = 0
   
   for other in my_list:
      if other == value:
         return i
         
      i += 1
    \end{lstlisting}
    \caption{Nested return statement example.}
    \label{fig:nest-ret-example}
    
    \end{subfigure}

    \caption{Python code segments inspired by student programs.}
\end{figure}

Consider the example in Figure \ref{fig:global-example} which contains code making use of a global variable.
Besides the fact that the use of a global variable is widely considered a bad practice, it is also in violation the principle of using explicit logic over implicit logic.
A rule to capture this is straight-forward,
\begin{align}\label{eq:global}
    \forall f, s\; Fun(f) \land Desc(f,s) \land Global(s) \implies M(f,s)
\end{align}

The predicate $Fun(f)$ determines if $f$ is a function, $Desc(f,s)$ ensures that $s$ is a descendent of $f$ in the AST, and $Global(s)$ is true if $s$ is a ``global'' statement.
If all three of these predicates are true, there is a mistake $M(f,s)$ in function $f$ in statement $s$ which is sufficient information to generate a helpful message.
The head, that is, the consequent of each rule in the model is simply a mistake predicate, hence no complex inference is necessary.
In general, the models' rules follow this pattern, the body of the rule consists of AST predicates and the head of the rule is a type of mistake.

In the previous example the type of mistake could easily be detected with a simple string search, however identifying the function they are found in is more difficult.
In general, the mistakes our model identifies requires knowledge of the AST.
Of particular importance is the ancestor-descendent relationship captured by the predicate $Desc$.
Consider the example found in Figure \ref{fig:nest-ret-example} which contains a segment of Python code with a nested ``return'' statement.
In this example, context is critical since ``return'' is legitimate and necessary in general.
However in this example, the code violates our principle of one-way-in-one-way-out as well as subtly contains a bug.
If the ``value'' is not contained in the list, the value ``None'' will implicitly be returned.
If the code calling the function always expects an integer return value the an exception will be raised.
Likewise this type of mistake is detected in a straight-forward fashion,
\begin{align}
    \forall f, s\; Fun(f) \land Desc(f,s) \land \lnot Child(f,s) \label{rule:nested} \\ \nonumber 
    \land\; Return(s) \implies M(f,s)
\end{align}

The predicate $Child(f,s)$ determines if $s$ is a direct child of $f$ and $Return(s)$ is true if $s$ is a return statement.
Here the subtly of the nestled ``return'' statement is readily captured by the use of the $Child$ and $Desc$ predicates.
A legitimate ``return'' will be directly under the function declaration in the AST, while a nested ``return'' will be deeper in the AST.
The contrast between $\lnot Child$ and $Desc$, that is a ``return'' statement being a syntactic descendent of a function declaration but not a direct child of it, is exactly the definition of a nested ``return''.

These two examples point out a key aspect of the model, the primary complex predicates needed are $Desc$ and $Child$, the rest of the predicates simply identify the type of expression or statement which is immediately apparent from the AST information.

\subsection{Implementation}

The models were implemented in the Haskell programming language.
While many languages could be a suitable choice, and some such as Prolog are even designed for logical deduction, Haskell has several natural advantages.
First, Haskell has support for parsing many common programming languages such as Python, Java, and JavaScript.
One benefit of Haskell is that the syntactic elements of Python and Java are expressed directly as Haskell data structures.
Functions in Haskell pattern-match on directly on data structures which means functions can easily be written to check specific parts of an AST.
For example, this means a function can be written to specifically check for errors in Java method declarations and associated statements.
Second, functional programming's declarative style and high-level functions mean the models' rules can be almost directly expressed in the language.
The logical rules of the model examine the AST for the existence or nonexistence of certain syntactic elements or identify particular relationships between them.
Most of these operations have a corresponding higher-order functions is used to implement them.

Because of differences in the Python and Java languages as well as subsequent differences in parsers, we decided to implement the Java and Python models as separate programs.
There are enough similarities that it is possible unify both programs, however it made more sense practically speaking to keep them separate.
Haskell is so effective at the task of representing our models, the implementation for Python is only 216 lines long while the Java implementation is 495 lines excluding comments and empty lines.



\subsection{Personalized Feedback}

The primary purpose of the system is to provide personalized feedback to students rapidly.
Our rule based model is fast, evaluating a student program in less than a second on a commodity computer.
The model was deployed to a web server and was made available to students with a simple HTML form.
There was nothing for the student to setup or install on their own computer, since the tutoring system was available as a web service.
Student could very easily use the system by simply submitting their code via the form.
The model was run immediately when a student submitted their program, producing feedback on design mistakes.
Since the model is fast, there was no need to queue up submissions or rate-limit the service at all.

Students were able to check their assignments for mistakes as many times as they wanted.
Further, the system was available 24/7, allowing the students to check their work whenever was convenient before the due date.
This meant the students had rapid feedback on their assignments and a chance to improve their work before being evaluated by the instructor.
Further, this increased the transparency of the grading process, since the rules are explicitly stated and the mistakes generated by them are easily understood.

\section{Experiments}

The effectiveness of the system was evaluated on student programs from multiple years of introduction to computer science courses.
The system was made available to students during both the first and second introduction to computer science courses.
Students were encouraged to use the system to evaluated their programs multiple times before submitting their program for grading.
This gave the students an opportunity to correct their mistakes before being graded.
The students were also informed that the output of the system was going to be used to a guide to instructors to grade their assignments, which provided incentive to use the system and transparency into the grading processing.

The experimental setup is simple, the Python programs from 2021 and the Java programs from 2022 are from students who had access to the tutoring system.
Submissions from previous years were used as a basis for comparison.
The actual assignments, standards, and requirements remained consistent over the years, the only difference between
experimental year and previous years is student access to and feedback from our system.
The system was used to quantify errors in the experimental and previous years' assignments.
This data was used to estimate the rate of mistakes made by students each year.
Hence the experiment compares the error rate of students who had access to the system versus students from previous years that did not.

\subsection{Dataset}

The dataset consists of programs from 4 student assignments collected over 4 years, 2 Python assignments taken from a introduction to computer science course and 2 Java assignments taken from the second semester introduction to computer science course.
All the programs included in the dataset were syntactically correct, all student submissions that could not be parsed into an AST simply were not included.
The Python assignments were the last 2 assignments of the course and were chosen because they were the most complex and lengthy assignments in the class.
The ``Craps'' assignment involves implementing the classic casino game as a command-line program.
Likewise the ``RPS'' assignment is the game of rock-paper-scissors.
The number of student programs totals to 506, with 109 of them being submitted in 2021.
Those submissions had our system available for feedback.
A summary of the Python data can be found in table \ref{table:pystats}.

The Java assignments are from the middle of the second semester introduction to computer science course.
The ``Car'' assignment involves creating a class to represent a Car, its fuel, MPG, and odometer as well as methods to make it ``drive'' etc.
Similarly, the ``Balloon`` assignment involves creating a class to represent a spherical balloon with its radius and has methods to inflate and deflate it with cubic units of air.
These methods require a mathematical conversion between radius and cubic units.
The total number of student submissions was 298 with 59 of them submitted in 2022.
A summary of the Java data can be found in table \ref{table:pystats}.
We did not record the number of times, if any, a student used our system to gain feedback.









\begin{table}[]
    \centering
    \begin{tabular}{|l|l|l|l|l|}
        \hline 
        \textbf{Program} & \textbf{Year} & \textbf{Programs} & \textbf{Mistakes} & \textbf{Rate} \\
        \hline
        \multirow{4}{*}{RPS (Python)} & 2018 & 68 & 311 & 4.57 \\
        &2019 & 74 & 285 & 3.85\\
        &2020 & 70 & 231 & 3.30\\
        &2021 & 55 & 145 & 2.64\\
        \hline
        \multirow{4}{*}{Craps (Python)} & 2018 & 65 & 388 & 6.0\\
        &2019 & 78 & 463 & 5.94\\
        &2020 & 42 & 210 & 5.00\\
        &2021 & 54 & 181 & 3.35\\
        \hline
        \multirow{2}{*}{Car (Java)} & 2020 - 2021 &  104 & 138 & 1.33 \\
        &2022 & 30 & 8 & 0.27\\
        
        \hline
        \multirow{2}{*}{Balloon (Java)} & 2020 - 2021 & 101 & 416 & 4.12\\
        &2022 & 29 & 74 & 2.55\\

        \hline
        
    \end{tabular}
    \caption{The statistics of student Python \& Java programs.
    Included are the number of student program submissions, the total number of design quality mistakes, and a rate of mistakes.
    Students only had access to the intelligent tutoring system during 2021 for Python and 2022 for Java. }
    \label{table:pystats}
\end{table}

\subsection{Results}

The rate of mistakes made by students significantly dropped with the introduction of our system.
For all the assignments, the drop in mistake rate was both statistically and substantial.
For the Python assignments, the comparison in mistake rates is between the 2021 programs and all previous years.
Similarly, for the Java assignments the comparison in mistake rates was between 2022 and the years 2020 and 2021.
The difference between the mistake rates was compared with a two-sample Poisson test.
For all assignments, the difference was found to be significant with p-values was less than .00002.
This is unsurprising considering the drop in the mistake rates was 32.31\%, 41.23\%, 79.70\%, \& 38.11\%  for ``RPS'', ``Craps'', ``Car'', and ``Balloon'' respectively.
The magnitude of the improvements clearly demonstrate the effectiveness of the system.
We believe that both the availability of the system along with the fact that it was used to guide assessment and grading provided the motivation to use the system and make corrections.
Further, the drop in design quality errors was noticeable for the instructors anecdotally.
Also, student appreciated the system as well, since it was fast, convenient, and provided additional transparency into the assessment process.

In the improvements in Java assignments were noticeable for two reasons.
First the dramatic nature of the improvement, as much as a 79.70\% drop in the rate of mistakes.
Second, was the fact that all the style guidelines had been written down specifically for Java and were available to students for years.
Unsurprisingly, students apparently did not read or apply the guidelines despite being told that their work would be assessed according to them.
However our tutoring system which directly encoded the guidelines, actually got the students to comply.
The difference is likely the dynamic nature of the system, giving precise, rapid feedback which is easier to utilize than reading an 8 page PDF of general rules.
It is likely far easier for students, especially relatively inexperienced ones, to respond to specific feedback over general statements.
Overall, the system was highly effective and that was only possible because of voluntary student participation.

\section{Conclusions \& Future Work}

Overall, an intelligent tutoring system was designed based to give rapid feedback on program design quality.
The model it implemented was derived from a set of design principles which were encoded as rules over abstract syntax trees for both Python and Java.
The system was fast, accurate, and was highly effective at delivering feedback to students.
As a result, the quality of student code improved substantially, also indicating their voluntary use of the system.

Though successful there are some types of design quality mistakes that the tutoring system does not catch.
Mistakes such as bad variable names, large blocks of repeated code, or uninformative comments cannot be identified by the system but would be a useful extension.
These kinds of mistakes are hard to typify directly with rules, so a machine learning model or other methods might be needed.



\medskip

\printbibliography


\appendix

\section{Python Model Rules}
\label{appendix:pyrules}

\begin{itemize}
    \item Global variables. \\ $\forall f,s \; Fun(f) \land Desc(f,s) \land Global(s) \implies M(f,s)$\\
    
    \item Break statements. \\ $\forall f,s \; Fun(f) \land Desc(f,s) \land Continue(s) \implies M(f,s)$\\
    
    \item Continue statements. \\$\forall f,s \; Fun(f) \land Desc(f,s) \land Pass(s) \implies M(f,s)$\\
    
    \item Pass statements. \\$\forall f,s \; Fun(f) \land Desc(f,s) \land Pass(s) \implies M(f,s)$\\
    
    \item Missing ``main'' function. \\$ (\forall f \; Fun(f) \land \lnot Name(f, ``main") ) \implies M'(``No\; Main")$\\
    
    \item Missing a call to ``main'' function. \\ $ \lnot (\forall f,s\; Fun(f)\; \land\; \lnot Desc(f,s)\; \land\; CallName(s, ``main")) \\ \implies M'(``No\;Call")$ \\
    
    \item ``Main'' function not first. \\$\forall f\; Fun(f) \land Name(f, ``main") \land \lnot First(f) \\
    \implies M'(``Not\; first")$\\
    
    \item ``Main`` function has arguments. \\$\forall f\; Fun(f) \land Name(f, ``main") \land HasArgs(f) \\
    \implies M'(``Main\; has\; Arguments")$\\
    
    \item No other function besides ``main''. \\ $ (\forall f\; Name(f, ``main")) \implies M'(``No\; other\; function")$ \\

    \item Nested function declaration. \\ $ \forall f,g \; Fun(f) \land Fun(g) \land Desc(f,g) \implies M(f,g) $ \\

    \item Nested ``return'' statement. \\ $\forall f,s \; Fun(f)\; \land\; Desc(f,s)\; \land\; \lnot Child(f,s)\; \land\; Return(s) \\
    S\implies M(f,s)$ \\

    \item Multiple ``return'' statements. \\ $\forall f,s,t\; Fun(f) \land Desc(f,s) \land Desc(f,t) \land Return(s) \land Return(t) \land s \ne t \implies M(f)$ \\

    \item Co-Recursive call to ``main''. \\ $\forall f,s\; Fun(f)\; \land\; Desc(f,s)\; \land\; \lnot\; Name(f, ``main")\;  \land\; Call(s)\; \land\; CallName(s, ``main")$ \\

    \item Recursive function call. \\ $\forall f,s\; Fun(f) \land Desc(f,s) \land Call(s) \land Name(f, s)\\
    \implies M(f,s)$ \\

    \item Calls to ``quit'' or ``exit'' functions. \\ $\forall s\; CallName(s, ``exit")\; \lor\; Call(s, ``quit")\\
    \implies M'(``Uses\; exit")$ \\

    \item Has magic numbers. \\ $ \forall f,e\; Fun(f) \land Desc(f,e) \land Magic(e) \implies M(f,e) $ \\

\end{itemize}

\section{Java Model Rules}
\label{appendix:jrules}

\begin{itemize}
    \item Attributes should be ``private'' or ``public static final''.\\
    $\forall c, a\; Class(c) \land Attribute(s) \land \lnot \big[ isPrivate(a) \\
    \lor\; \big( isPublic(a)\; \land\; isStatic(a)\; \land\; isFinal(a) \big) \big] \\
    \implies M(c,a)$\\
    
    \item Attribute name should have a preceding underscore.\\
    $\forall c, a\; Class(c) \land Attribute(a) \land \lnot GoodPrefix(a) \\ \implies M(c,a)$\\
    
    \item Final attribute's name should be in all-caps.\\
    $\forall c, a\; Class(c)\; \land\; Attribute(a)\; \land\; \lnot AllCaps(a) \\ \implies M(c,a)$\\
    
    \item One attribute declaration per line (block).\\
    $\forall c, d, a, a^\prime\; Class(c)\; \land\; DeclarationBlock(c, d) \\
    \land \; InBlock(d,a) \land \; InBlock(d,a^\prime) \land \; a \ne a^\prime \\
    \implies M(c, a)$\\
    
    \item No initializer block.\\
    $\forall c, b\; Class(c)\; \land \; InitializerBlock(b) \implies M(c,b)$\\
    
    \item Has magic numbers, same as Python model.\\
    
    \item Multiple ``return'' statements, same as Python model.\\
    
    \item Break statements, same as Python model.\\
    
    \item Continue statements, same as Python model.\\
    
    \item Methods are limited to 30 statements.\\
    $\forall m\; Method(m)\; \land\; DescCount(m, 30) \implies M(m)$\\
    
    \item ``instanceof'' operator.\\
    $\forall m, s\; Method(m)\; \land\; Desc(m,s)\; \land\; isInstanceOf(m,s)\\
    \implies M(m,s)$\\
    
    \item Ternary operator.\\
    $\forall m, s\; Method(m)\; \land\; Desc(m,s)\; \land\; isTernary(m,s)\\
    \implies M(m,s)$\\
    
    \item Labeled statement.\\
    $\forall m, s\; Method(m)\; \land\; Desc(m,s)\; \land\; isLabel(m,s)\\
    \implies M(m,s)$\\
    
    \item Lambdas expression.\\
    $\forall m, s\; Method(m)\; \land\; Desc(m,s)\; \land\; isLambda(m,s)\\
    \implies M(m,s)$\\
    
    \item On-the-fly local variable declaration.\\
    $\forall m, x, y \; Method(m)\; \land Desc(m,x)\; \land Desc(m,y)\; \land isVarDecl(x)\; \land\; \lnot isVarDecl(y) \; \land \; Before(y, x) \\ \implies M(m, x) $\\
    
    \item ``if'' statement has block.\\
    $\forall m, s, t\; Method(m)\; \land\; Desc(m,s)\; \land Child(s,t)\; \land isIf(s)\; \\
    \land \lnot IsStmtBlock(t) \implies M(m, s) $\\
    
    \item ``while'' loop has block.\\
    $\forall m, s, t\; Method(m)\; \land\; Desc(m,s)\; \land Child(s,t)\; \land isWhile(s)\; \\
    \land \lnot IsStmtBlock(t) \implies M(m, s) $\\
    
    \item ``for'' loop has block.\\
    $\forall m, s, t\; Method(m)\; \land\; Desc(m,s)\; \land Child(s,t)\; \land isFor(s)\; \\
    \land \lnot IsStmtBlock(t) \implies M(m, s) $\\
    
    \item C-style ``for'' loop is conventional.\\
    $\forall\; m, f, s, c, i \; Method(m)\; \land\; isFor(f)\; 
    \land\; ifForInit(f,s)\; \land\; isForCond(f,c)\; \land\; isForInc(f,i)\; 
    \land\; isVarDecl(s)\; \land\; isBinaryOp(c)\; \land\; 
    isUnaryOp(i) \implies M(m, f)$\\
    
    \item One local variable declared per line.\\
    $\forall m, d, a, a^\prime\; Method(c)\; \land\; DeclarationBlock(c, d)\;
    \land \; InDeclBlock(d,a) \land \; InDeclBlock(d,a^\prime) \land \; a \ne a^\prime \\
    \implies M(m, a)$\\
    
\end{itemize}

\end{document}